\documentstyle[sprocl,epsf]{article}

\begin{document}

\def\bea{\begin{eqnarray}}
\def\eea{\end{eqnarray}}
\newcommand{\beq}{\begin{equation}}
\newcommand{\eeq}{\end{equation}}

\title{Polchinski ERG Equation in $O(N)$ Scalar Field Theory \footnote{
Talk given by Rui Neves at the Second Conference on the Exact Renormalization Group\\ 
\hbox{}\hspace{0.6cm}(Rome, September 18-22, 2000).}}
\author{Yuri Kubyshin\\
\vspace{0.1cm}
\footnotesize{Institute for Nuclear Physics, Moscow State University\\
119899 Moscow, Russia}\\
\vspace{0.1cm}
Rui Neves and Robertus Potting\\
\vspace{0.1cm}
\footnotesize{\'Area Departamental de F\'{\i}sica, FCT, Universidade do Algarve\\
Campus de Gambelas, 8000 Faro}}

\maketitle

\begin{abstract}
We investigate the Polchinski ERG equation for $d$-dimensional 
$O(N)$ scalar field theory. In the context of the non-perturbative derivative
expansion we find families of regular solutions and establish their relation
with the physical fixed points of the theory. Special emphasis is given to the
limit $N=\infty$ for which many properties can be studied analytically.  
\end{abstract}

\section{Introduction}

Over the years the exact renormalization group (ERG)
\cite{KW,SW,WH,JP} has grown to become a reliable and accurate framework 
in the study of non-perturbative phenomena in quantum field theory (see reviews
for example in \cite{TM1,YK,JW,BB}).

In this context the Polchinski ERG approach and the derivative expansion 
\cite{JP,BT,TM2,BHLM} are specially attractive for their power and
simplicity. These qualities are well in evidence in the study of an $N=1$
scalar field theory \cite{BHLM,YRR}. Going a step further we now consider the
$d$-dimensional $O(N)$ models \cite{CT,TM3}. Their Polchinski ERG equation 
in the local potential approximation has been analyzed by Comellas and 
Travesset \cite{CT}. Namely, fixed-point solutions and the corresponding
critical exponents were determined and the large $N$
limit discussed in some detail. 

In the leading order of the derivative expansion the field
renormalization is neglected and the anomalous dimension $\eta$ is
set to zero. As a consequence, features for which the field renormalization
is essential are lost. An example is
the $N=1$ scalar field theory in two dimensions. If $\eta=0$ then only the
continuum limits described by periodic solutions and corresponding to the
critical sine-Gordon models are seen in the ERG approach \cite{YRR,TM}.
However, it was shown \cite{YRR} that the set
of non-perturbative 2D conformal fixed points, found in the second order of the
derivative expansion \cite{TM}, is already seen in the space of regular
solutions of the leading order Polchinski equation with non-zero $\eta$. 
Furthermore, the analysis was shown to be valid for any dimension
$d$ though not all regular solutions corresponded to physical fixed points
\cite{YRR}. This motivates the present investigation where we study the space
of solutions of the leading order Polchinski equation in $d$-dimensional
$O(N)$ scalar field theory. We expect that this enables us to gain an insight
into a better understanding of the Polchinski ERG approach and 
provide examples of leading order solutions that are needed for the analysis 
of higher orders in the derivative expansion. In this work we focus on the
presentation of our results leaving a more detailed technical description for
a forthcoming publication \cite{YRR1}. 

In Section 2 we study the $O(N)$ scalar field theory in the limit
$N=\infty$ where many properties can be deduced analytically. In
Section 3 we consider some aspects of the $O(N)$ models with finite $N$. In
Section 4 we present our conclusions.   

\section{$O(N)$ scalar field theory: the $N=\infty$ limit}              

\begin{figure}[t]
\epsfxsize=11.75cm
\epsfbox{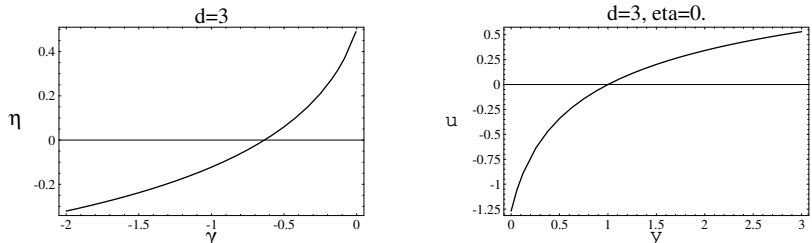}
\vspace{-1.5cm}
\caption{Plots of the functions $\eta_{1} (\gamma)$ and 
${u_1}(y;{\gamma_*})$, ${\gamma_*}=-0.634915\ldots$, for $d=3$.   
\label{fig:eg0uy3}}
\end{figure}

For large $N$ the leading order Polchinski ERG fixed-point 
equation can be written as follows

\beq
- 2yuu' - u^{2} + 
\left( 1 + 2y\Delta^{-} \right) u' + su = 0,\quad u(0) = 2\gamma. \label{P1}
\eeq
Here we introduced the notations ${\Delta^\pm}=1\pm d/2-\eta/2$,  
$s={\Delta^+}+{\Delta^-}=2-\eta$, $2y\equiv \sum_{i=1}^{N} \varphi_{i} \varphi_{i}$, where $\varphi_i$ are the components of the $O(N)$ field, and $u(y)=v'(y)$ is the
derivative  with
respect to $y$ of the potential $v$ in the Wilsonian effective action.

Two obvious solutions of Eq. (\ref{P1}) are $u(y)=0$ and $u(y)=s$
which correspond to the Gaussian (GFP) and to the trivial (TFP)
fixed points respectively. To find non-trivial solutions we follow the analysis of
Comellas and Travesset \cite{CT} and consider the fixed-point equation for the
inverse function $y(u)$

\beq
(u-s)uy'-2\left({\Delta^-}-u\right)y-1=0,\quad y(2\gamma)=0,
\label{ifpe}
\eeq
where the prime now represents the derivative with
respect to $u$.

The general solution of Eq. (\ref{ifpe}) is given by
\beq
y(u) = \frac{1}{(s-u)^{2-\alpha} u^{\alpha}} \left[ 
C(\eta,\gamma) - \frac{u^{\alpha} (s-u)^{1-\alpha}}{\alpha} 
+ \frac{\alpha - 1}{\alpha} {\int_0^u} 
\left( \frac{s - z}{z} \right)^{-\alpha} dz \right], \label{yu1}
\eeq
where $\alpha=2{\Delta^-}/s$ and the integration constant $C(\eta,\gamma)$ is 
equal to 
\beq
C(\eta,\gamma) =
{{{{(2\gamma)}^\alpha}{{(s-2\gamma)}^{1-\alpha}}}\over{\alpha}}-
{{\alpha-1}\over{\alpha}}{\int_0^{2\gamma}}{{\left({{s-z}\over{z}}
\right)}^{-\alpha}}.\label{icons}
\eeq
In order that $y(u)$ be analytic we need to impose $-1<\alpha<0$. 
Let us consider a point $y_{*}$ such that
$2{y_*}{u(y_*)}-(1+2{y_*}{\Delta^-})=0$  with $u(y_{*}) \neq 0$. It can be
checked that the solution $u(y)$ is non-analytic at $y={y_*}$, namely $u'(y)$
is divergent at $y={y_*}$. For a solution to be regular we impose the
additional condition $u(y_{0}) = 0$ (or $y(0) = {y_0}$), where 
$y_{0}=- 1 / (\alpha s)$. This gives certain constraints on the parameters of
the solution. We distinguish the following two classes of solutions.

1) For $C(\eta,\gamma)=0$ we get a linear 
behaviour for $u(y)$ in the vicinity of $y_0$,
\beq
u(y) = - \frac{\alpha (1+\alpha) s^{2}}{2}  (y - y_{0}) + \cdots . 
\label{uy-a}
\eeq
Let us label this type of regular solutions $u(y)$ by $n=1$. They are 
calculated by inverting the function $y(u)$ in Eq.
(\ref{yu1}) with $C(\eta,\gamma)=0$. 

The condition $C(\eta,\gamma)=0$ can be written as 
\beq
-\frac{2\gamma}{d} (1-\alpha)^{2} 
\left(1 - \frac{2\gamma}{d} (1-\alpha) \right)^{\alpha -1} \int_{0}^{1} dz 
z^{\alpha} \left( 1- \frac{2\gamma}{d} (1-\alpha) z \right)^{-\alpha} = 1.  
\label{C-0}
\eeq
Eq. (\ref{C-0}) defines $\alpha_{1}$ as a function of $\gamma/d$ for 
$\gamma < 0$. Correspondingly, the anomalous dimension 
$\eta_{1} (\gamma) = 2 - d/(1-{\alpha_1}(\gamma /d))$. 
For $\gamma < 0$ and $\gamma \rightarrow {0^-}$ we find that $\alpha_{1}
(\gamma/d) \to -1$ and therefore $\eta_{1}(\gamma) \to 2-d/2$. This
corresponds to the GFP. For $\gamma \to -\infty$ the function $\alpha_{1}
(\gamma/d) \to 0$ and, correspondingly, $\eta_{1}(\gamma) \to 2-d$. 

To obtain the curve $\eta = \eta_{1}(\gamma)$ we
solved Eq. (\ref{C-0}) numerically. For $d=3$ the result is given in
Fig.\,\ref{fig:eg0uy3}. In other dimensions the function
$\eta_{1}(\gamma)$ has a similar profile and can be easily computed
from the curve for $d=3$ using the scaling properties following from Eq. 
(\ref{C-0}) and the definition of $\eta_{1}(\gamma)$. Each
point on the curve ${\eta_1}(\gamma)$ corresponds to a regular solution
which we denote as $u_{1}(y;\gamma)$. For $d=3$ the
curve ${\eta_1}(\gamma)$ crosses the $\gamma$-axes at 
$\gamma = \gamma_{*} = - 0.634915 \ldots$ and therefore 
it corresponds to a physical fixed point. In the vicinity of $y_0$ the
solution has the linear behaviour (\ref{uy-a}) that indicates that 
it corresponds to the Heisenberg fixed point (HFP) \cite{CT}. 
The function $u_{1}(y;\gamma_{*})$ is given in Fig.\,\ref{fig:eg0uy3}. 
The numerical results for the curve 
${\eta_1}(\gamma)$ show that
non-trivial regular solutions exist only for $\gamma<0$ and $-1<\eta<
1/2$. For $\gamma=0$, $\eta=1/2$ we have the GFP. 

\begin{figure}[t]
\epsfxsize=11.75cm
\epsfbox{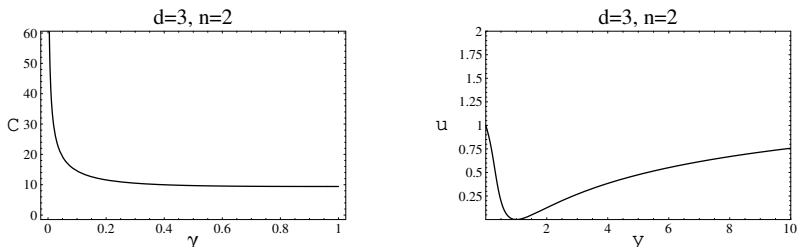}
\vspace{-1.5cm}
\caption{Plots of the functions ${C_2}(\gamma)$ and 
${u_2}(y;{\gamma})$, ${\gamma}=1/2$, for $d=3$.
\label{fig:cguy32}}
\end{figure}

2) For $C(\eta,\gamma) \neq 0$ we find   
\vspace{-0.2cm}
\beq
u(y) = s^{-\frac{2-\alpha}{\alpha}} \left[ \frac{y - y_{0}}{C(\eta,\gamma)} 
\right]^{-\frac{1}{\alpha}} + \cdots    \label{uy-b}
\eeq
The condition of regularity of the solution $u(y)$ at $y=y_{0}$ gives 
$(-1/\alpha) \equiv - 2{\Delta^-}/s = n = 2,3,\ldots$. Hence, solutions from
the second class are labelled by integers $n\geq 2$.  They have the power-law
behaviour (\ref{uy-b}) in the vicinity  of the potential singularity $y_{0}$
and are obtained by inverting  the exact formula (\ref{yu1}) with 
$\eta = \eta_{n}$ given by ${\eta_n} = 2 - dn/(n+1)$.

Let us denote the function (\ref{icons}) with $\eta = \eta_{n}$ by 
$C_{n}(\gamma)$. From its definition it follows that for $n$ even the function
$C_{n}(\gamma)$ is analytic for $0 < 2\gamma \leq s_{n}$,  
$s_{n} = 2 - \eta_{n}$.
These general features are illustrated by the the plot of the function
${C_2}(\gamma)$ for $d=3$ presented in Fig.\,\ref{fig:cguy32}. For each
finite value of ${C_n}(\gamma)$, $n$ even, there is a regular solution 
$u_{n}(y;\gamma)$. It is obtained by inversion of the function (\ref{yu1}) 
with $C(\eta,\gamma)=-{C_n}(\gamma)$ for $0\leq y\leq{y_0}$ and with
$C(\eta,\gamma)={C_n}(\gamma)$ for $y\geq{y_0}$. In Fig.\,\ref{fig:cguy32} we
show the plot of the function ${u_n}(y;\gamma)$ for $d=3$, $n=2$ and
$\gamma=1/2$.

\begin{figure}[t]
\epsfxsize=11.75cm
\epsfbox{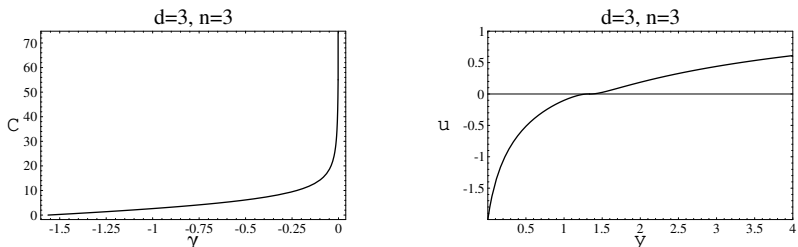}
\vspace{-1.5cm}
\caption{Plots of the functions ${C_3}(\gamma)$ and 
${u_3}(y;{\gamma})$, ${\gamma}=-1$, for $d=3$.   
\label{fig:cguy33}}
\end{figure}

One can easily see that as $\gamma \rightarrow 0$ the function  
$C_{n}(\gamma) \rightarrow +\infty$. This limit, of course, corresponds to the
GFP. For $\gamma = s_{n}/2$ we find 
\[
C_{n}(s_{n}/2) = nd \frac{\pi \alpha_{n}}{\sin (\pi \alpha_{n})} = 
\frac{\pi d}{\sin \frac{\pi}{n} }.
\]
Correspondingly, the solution is analytic at $u=s$ with
$y(s)=1/(2{\Delta^+})$ and so it
does not satisfy the initial condition $u(0)=2\gamma$. Moreover, the 
behaviour for $y\to 0$ is non-analytic, namely $u \sim 1/y$.
This intriguing fixed point \cite{CT} is clearly distinct from those for which
$0<\gamma<s/2$. For $d=3$ ${\eta_2}=0$ and the corresponding solution is
physical. Further analysis of its properties is beyond the 
scope of the present article.

For $n$ odd $C_{n}(\gamma)$ must be positive or zero, therefore it
exists only for ${\gamma_{*n}}\leq\gamma < 0$, where
${C_n}({\gamma_{*n}})=0$. In Fig.\,\ref{fig:cguy33} we show the 
plot of the function $C_{3}(\gamma)$ 
for $d=3$. The regular solution, corresponding to 
$\gamma=-1$ is also given in Fig.\,\ref{fig:cguy33}. Note that
when ${C_n}(\gamma)=0$ we reach the HFP.

\section{$O(N)$ scalar field theory: finite N}

For arbitrary but finite $N\geq 1$ the leading order Polchinski 
fixed-point equation is given by

\beq
{2y\over N}u''+(1+{2\over
N}+2\Delta^- y-2yu)u'+(\Delta^+ +\Delta^-
-u)u=0, \quad u(0)=2\gamma.    \label{eq:uN}
\eeq

Solving it numerically we found that the non-trivial regular solutions
correspond to points of the curves $\eta_{n}(\gamma)$, $n\geq 1$. For $n$ even,
the curves lie in the region $\gamma>0$ bounded by the lines $\gamma=0$ (GFP),
$\eta=2-d$ and $\gamma=1-\eta/2$ (TFP). For $n$ odd the region is $\gamma<0$ 
bounded by the lines $\gamma=0$ and $\eta=2-d$. Higher values of $n$ correspond
to lower-lying curves (see Fig.\,1 of Ref. \cite{YRR} for an ilustrative example).  

\begin{figure}[t]
\epsfxsize=11.75cm
\epsfysize=5cm
\epsfbox{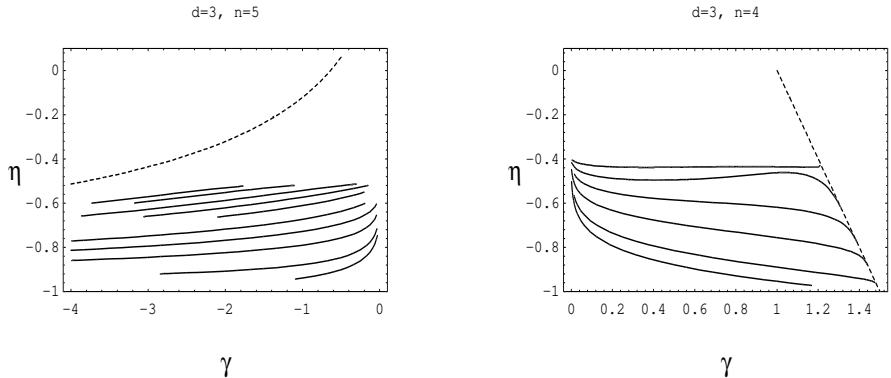}
\caption{Curves corresponding to $d=3$, $n=5$ ($N=1$, 2, 5, 10, 20,
50, 100, 200, 500, 1000) and $n=4$ ($N=1$, 2, 5, 10, 20, 50).
\label{fig:testN}}
\end{figure}

It is also interesting to compare the curves that correspond to a fixed
value for $n$ but for a range of values of $N$.
In Fig.\,\ref{fig:testN} we have plotted the case $n=4$ for $d=3$ where the
TFP line is also included.
We see that the larger the value of $N$ is, the higher the corresponding curve
lies. For large values of $N$ they approach the horizontal line $\eta=\eta_n$
for $0<\gamma<s_n/2$ and the TFP line for $s_n/2 <
\gamma < 3/2$. 
The other plot in Fig.\,\ref{fig:testN} shows the case $n=5$ for
$d=3$. Again, larger values of $N$ correspond to higher lying curves.
Here we see that the curves
tend to the $N=\infty$ horizontal line $\eta = \eta_{5}$ for
$\gamma_{*5}<\gamma<0$ and to the $C=0$ curve $\eta = \eta_{1}(\gamma)$ for
values $\gamma<\gamma_{*5}$.

\section{Conclusions}

In this work we have studied the solutions of the leading order Polchinski
ERG fixed-point equation for the $O(N)$ scalar field theory. 
We have described the space of regular solutions corresponding to points of
the curves $\eta_{n}(\gamma)$ in the $(\gamma,\eta)$-plane. If for a given
$d$ the curve $\eta_{n}$ crosses the $\gamma$-axis at $\gamma=\gamma_{*}$,
then the corresponding solution is the physical fixed-point solution with
$u_{n}(0;\gamma_{*})=2\gamma_{*}$. We found that the pattern of the curves of
regular solutions is universal for any $d$. In this sense each curve
represents a fixed-point solution of a certain type.     

We have paid special attention to the limit $N=\infty$ where many
properties of the space of solutions can be investigated analytically. A
general solution is non-analytical at some finite point and, therefore, is
not acceptable from the physical point of view. Analyticity
imposes a relation between $\eta$ and $\gamma$, thus giving rise to the curves
$\eta_{n}(\gamma)$ in the $(\gamma,\eta)$-plane labelled by an integer $n=1,2,
\ldots$. For $n \geq 2$ the curves are in fact horizontal straight lines. 

For the $N\geq 1$ field theory we also obtained continuous non-trivial 
fixed-point lines labelled by an integer $n \geq 1$. Clearly, there is a
one-to-one correspondence between the lines with equal value of $n$ for $N\geq
1$ and for $N=\infty$. We also found a strong indication that the functions
$\eta_{n}(\gamma)$ for finite $N$ transform into the corresponding $N=\infty$
functions $\eta_{n}(\gamma)$ as $N \to\infty$.

\section*{Acknowledgements}

We acknowledge financial support from the Portuguese Funda\c {c}\~ao
para a Ci\^en\-cia e a Tecnologia under grant number
CERN/P/FIS/15196/1999. R.N. also acknowledges financial support from 
fellowship PRAXIS XXI/BPD/14137/97.

\end{document}